\documentclass[a4paper, floatfix]{revtex4} 
\usepackage{vmargin,color} 
\usepackage[french,english]{babel} 
\usepackage[utf8]{inputenc} 
\usepackage{graphicx,float,setspace,amssymb,amsmath} 

\newcommand{\equa}[1]{\begin{eqnarray} \label{#1}} 
\newcommand{\auqe}{\end{eqnarray}} 
\newcommand{\tab}[1]{\begin{tabular}{#1}} 
\newcommand{\bat}{\end{tabular} \\ } 
\newcommand{\blanc}{\makebox[1.1 cm]{ }}

\newcommand{\al}{\alpha}  
 \newcommand{\ep}{\epsilon} 
\newcommand{\de}{\delta} \newcommand{\gD}{\Delta}  
\newcommand{\gf}{\varphi} \newcommand{\gl}{\lambda}
 
\newcommand{\tend}{\rightarrow} 
\newcommand{\En}{E[\hat{m}(z)]}
\newcommand{\dd}[2]{\frac{\partial #1 }{\partial #2 }}

\newcommand{\fand}{\vphantom{\left[ \left[ A^2 \right]^2 \right]}}
\providecommand{\abs}[1]{\left\vert#1\right\vert} 
 
\begin{document} 
\selectlanguage{english} 
\title 
{{\bf 
Nanostructured exchange coupled hard~/~soft composites:
from the local magnetization profile to an extended 3D simple model 
}} 
\author 
{ V. Russier, K. Younsi and L. Bessais \\
ICMPE, UMR 7182 CNRS and University UPEC, \\
2 rue Henri Dunant, 94320 Thiais, France.
} 
%
\thispagestyle{empty} 
\begin{abstract} 
In nanocomposite magnetic materials the exchange coupling between phases plays a central role
in the determination of the extrinsic magnetic properties of the material: coercive field,
remanence magnetization. Exchange coupling is therefore of crucial importance in composite 
systems made of magnetically hard and soft grains or in partially crystallized media including 
nanosized crystallites in a soft matrix. 
It has been shown also to be a key point in the control of stratified hard / soft media 
coercive field in the research for optimized recording media. 
A signature of the exchange coupling due to the nanostructure
is generally obtained on the magnetization curve $M(H)$ with a plateau characteristic  
of the domain wall compression at the hard/soft interface ending at the depinning
of the wall inside the hard phase. This compression / depinning behavior
is clearly evidenced through one dimensional description of the interface, which is
rigorously possible only in stratified media. Starting from a local description of the  
hard/soft interface in a model for nanocomposite system we show that one can extend
this kind of behavior for system of hard crystallites embedded in a soft matrix.
\end{abstract} 
\maketitle
\section {Introduction} 
\label{intro} 

A simple model in order to understand the relation between the nanoscale structure and the 
extrinsic magnetization curves in partly crystallized rare earth transition metal alloys (Re-M), 
can be built as an assembly of nanosized crystallites embedded in an amorphous matrix. 
Such a model is based upon experimental caracterization of
the samples obtained from an amorphous precursior, obtained either by high energy milling or 
a melt spun technique for instance, followed by an appropriate 
annealing which induces the re-crystallization 
\cite{hadjipanayis_1999,chen_1999,chen_2000,younsi_2010}.
The crystallites and the matrix must be characterized by hard and soft magnetocrystalline 
anisotropy respectively. One key point concerning the extrinsic magnetic properties of the 
system is then the exchange coupling between the crystallites and the matrix or between the 
crystallites {\it via} the matrix.
In other words, since one deals with nanocomposite including magnetic hard and soft phases, 
such models can enter in the so-called exchange spring magnet (ES) systems
\cite{kneller_1991,
fullerton_1998,victora_2005,suess_2006,suess_2007,goncharov_2007,skomski_2008}.
Exchange coupling between different phases of nanostructred composite materials have been widely
studied after the pioneering work of Aharoni \cite{aharoni_1960} 
mainly through the behavior of the average magnetization curve in terms on the
applied field, $M(H)$ \cite{hadjipanayis_1999,asti_2004,asti_2006,dobin_2006,suess_2007,ghidini_2007,pellicelli_2010}. 
On the other hand the nucleation field corresponding to the very beginning
of the magnetization reversal in the soft phase has been determined semi-analytically in different
situations from the linearization of the micromagnetic energy with respect to the deviation of
the local magnetization from the direction imposed by the hard phase easy axis
\cite{skomski_1993,skomski_1994,asti_2006,pellicelli_2010}.
Conversely, the local magnetization {\bf m}({\bf r}) in the vicinity of the interface between
phases in nanocomposites has not been investigated in details but in continuous film geometry
(layered systems) \cite{leineweber_1997,kronmuller_2002,guslienko_2004},
where furthermore the effect of the hard and soft layers thickness has been 
investigated through the magnetic phase diagram representing the reversal mode of the 
demagnetization, namely rigid mode versus ES-coupled mode,
\cite{asti_2004,asti_2006,ghidini_2007,pellicelli_2010} {\it via} the analytical 
determination of the magnetic susceptibility.
In the works devoted to the magnetization reversal in ES layered systems through the 
behavior of the local magnetization in terms of the distance, $z$, normal to the interface 
the solvation of the Euler equations relative to the micromagnetic energy minimization is 
made tractable because of the symmetry parallel to the surface according to which the local 
magnetization depend only on $z$ \cite{leineweber_1997,kronmuller_2002,guslienko_2004}.
This makes the nucleation, domain wall (DW) compression and depinning process at the 
hard~/~soft interface a well established theoretical result 
\cite{fullerton_1998,kronmuller_2002,dobin_2006,suess_2006}
which moreover has been experimentally observed \cite{fullerton_1998,casoli_2008,tsai_2010}.
In the case of nanocomposites made of true 3D grains either simply juxtaposed or separated by
grain boundaries, the situation is not so clear and as a general rule one refers to the 
similarity in the $M(H)$ curve to map the magnetization reversal process to the one indeed 
obtained for the ES layered system.

In the present work, we deal with a model made of hard inclusions embedded in
a soft matrix with the purpose to exploit the above mentioned theoretical results on
the magnetization reversal in magnetically hard/soft composite systems.
More precisely our aim is to bridge the gap between the stratified media, for which the 
reversal process is well understood, and our system of magnetically hard material inclusions
embedded in the soft matrix. We investigate the way according to which one can map the well-known 
nucleation, DW compression and depinning process of the magnetization reversal characteristic 
of the exchange coupled stratified media to the exchange coupled inclusion~/~matrix system. 
We will use both analytical results for the angle profile at the
planar interface in a one dimensional approximation and numerical results for true 3-D 
systems obtained from a micromagnetism finite element based code (MAGPAR) \cite{magpar_2003}.

In a first step, we compare the 1-D profile to the simulated result for one cubic inclusion 
embedded in a prismatic matrix. 
The easy axis of the inclusion is parallel to the interface (parallel recording media type of geometry).
The variation of $M(H)$ in the plateau region characteristic of layered ES media
is very well reproduced, and the accuracy obtained for the magnetization
profile, shows that the nucleation, domain wall compression and depinning process is valid for 
isolated inclusions embedded in a very soft matrix.

In order to investigate the effect of the finite value taken by the edge to edge separation 
$\gD$ between the crystallites in the actual system, we consider a model made of 
two cubic inclusions in the soft matrix. 
The magnetization profile behavior in terms of the applied field shows that the magnetization 
reversal follows the nucleation, domain wall compression and depinning process when $\gD$ 
is larger than the exchange length $l_{ex}$ of the soft material whereas when $\gD$ gets 
smaller than $l_{ex}$ the magnetization reverses in a coherent way in the whole system 
which corresponds to the strong coupling regime, or the rigid composite magnet of Refs. 
\cite{asti_2004,asti_2006}, where the system behaves as a single phase.

Then, we consider the effect of the misallignment between the easy axes of the neighboring 
inclusions. In this case, the domain wall which was of Bloch type for the parallel oriented 
interface, takes clearly a N\'eel character, with nevertheless at the qualitative level, the 
same type of magnetization reversal processes.

\section {One dimensional model }
\label   {1d_model}

In this section, we first recall the basic equations concerning the one dimensional 
description for the magnetization profile accross the hard soft interface.
We consider the planar sharp interface between a uniaxial hard material and an ultra soft
material. The interface, normal to the $\hat{z}$ axis, is defined by $z = z_0$
and the easy axis of the hard 
material, this latter being located at $z > z_0$, coincides with the $\hat{x}$ axis. 
The micromagnetic characteristics of the media will be denoted $K_{s,h}, A_{s,h}, J_{s,h}$
where the subscript $s (h)$ refers to the soft (hard) phase and $K, A, J$ denote 
the magnetocrystalline anisotropy constant, the exchange stiffness and $\mu_0$ times 
the saturation magnetization respectively. 
Here we shall consider the limiting case of an ultra soft phase, with $K_s = 0$.
Then the properties of the one dimensional model are totally determined by the
reduced parameters $\ep_A = A_s/A_h$, $\ep_J = J_s/J_h$, 
the reduced distance $z* = z/\de_B$, $\de_B$ being the Bloch DW thickness in the 
hard phase, $\de_B = \pi\sqrt{A_h/K_h}$ and
the reduced field $h = -H_a/H_K$
where $H_a$ and $H_K = 2K_h/J_h$ are the applied and the hard-phase anisotropy field respectively. 
The minus sign in the definition of $h$ is only for convenience in order to deal 
with $h > 0$ in the second quadrant $H_a < 0$.
We consider the case of an external field ${\bf H}_a$ in the $\hat{x}$ direction.
Starting from the energy functional where we explicitly assume the magnetization profile 
to lie in the plane $(\hat{x}, \hat{y})$ and to depend only on $z$
\equa{ener_1}
\En &=& S \int \left[ 
A(z)\left( \left( \dd{m_x(z)}{z} \right)^2 + \left( \dd{m_y(z)}{z} \right)^2 \right)
  +  K(z)(1 - m_x(z)^2) - J(z) m_x(z)H_a  \right] dz  \nonumber \\
X(z) & = & X_s\Theta(z_0 - z) + X_h\Theta(z - z_0); \blanc X = K, A \textrm {  or } J.
\auqe
where $\Theta(z)$ is the Heavyside step function. The minimization of the functional $\En$ 
with respect to $\hat{m}(z)$ leads to the well known Euler equations 
\cite{leineweber_1997,kronmuller_2002,guslienko_2004,asti_2004,asti_2006}
which, written in terms of the angular profile $\gf(z) = (\hat{m}(z), \hat{x})$ reads
\equa{phi_1}
\dd{\gf(z)}{z} &=& \pm \frac{1}{\sqrt{A_{h,s}}} \left[ \gD U_{h,s} (\gf(z))  \right]^{1/2} \nonumber \\ 
  \textrm{ with } \gD U_{h,s} (\gf(z)) &=& K(z) \left( \sin^2(\gf(z)) - \sin^2(\gf(z_b^{(h,s)})) \right)
- J(z) H_a \left( \cos(\gf(z)) - \cos(\gf(z_b^{(h,s)})) \right)
\nonumber \\
\auqe
The singularity of the interface leads to the boundary condition at $z = z_0$
\equa{cond_lim_z0}
  \left. \ep_A~ \dd{\gf}{z} \right]_{z_0^-} = \left. \dd{\gf}{z} \right]_{z_0^+}
\auqe
In equation (\ref{phi_1}) the choice for the $\pm$ sign depends on the boundary conditions far 
from the interface, represented by the values taken by $\gf_b^{(s,h)}~=~\gf(z_b^{s,h})$
where $z_b^{(s,h)}$ denote the location of either the soft or the hard layer center. 
In the case of the hard / soft interface with $z_b^{(s,h)}~\tend~\pm\infty$ we shall
asume in equation (\ref{phi_1}) that far from the interface, $z \sim z_b$, 
the magnetization 
$\hat{m}$ is alligned to the easy axis and that either $K_{h,s} > 0$ or $K_h > 0$ and $K_s = 0$
($\sin\gf_b = 0$).
Starting from an external field applied in the $x > 0$ direction, where $\hat{m}(z) = \hat{x}$
in the whole system, the field is decreased and then the first step in the magnetization
reversal is the so-called nucleation field $h = h_{nucl}$ corresponding to the reversal in the 
soft phase. $h_{nucl}$ can be obtained from the expansion of $\En$ in terms of the small deviations 
$\de \vec{m} = (\hat{m}(z) - \hat{x})$ at second order in $\abs{\de \vec{m}}$ \cite{skomski_1993}. 
In the following we limit ourselves to a hard layer thick enough for 
$z_b^{h}$ to be considered as infinite. Let us first consider the case $z_b^{(s)} \tend \infty$.
As usual, we consider the situation where the magnetization 
in the soft phase, far from the interface is along the field, namely 
$\hat{m}(z\sim z_b^{(s)}) = - \hat{x}$.
The profile  $\gf(z)$ can be easyly obtained from the numerical inversion of
\equa{profil_1}
  z - z_0 = \int_{\gf_0}^{\gf(z)} \dd{z}{\gf} d\gf = 
  -\int_{\gf_0}^{\gf(z)} \sqrt{A(z)}\left[ \gD U_{h,s}(\gf) \right]^{-1/2} d\gf
\auqe
which can be integrated, with the result 
\equa{profil_2}
  z^* - z^*_0 &=&    \left. \frac{\sqrt{\ep_A/\ep_J}}{\pi} 
               \ln \left(\frac{1 - \tan(\gf/4)}{1 + \tan(\gf/4)} \right) \right]  _{\gf_0}^{\gf(z)} ;
   \textrm{ for } \gf > \gf_0 \nonumber \\
  z^* - z^*_0 &=& -\frac{1}{2\pi\sqrt{1-h}} \left. 
               \ln \left( \frac{\gD - \cos(\gf/2) }{\gD + \cos(\gf/2) }\right) \right]  _{\gf_0}^{\gf(z)} 
   \textrm{ with } \gD = \sqrt{1 - \frac{\sin^2(\gf/2)}{1 - h}} ; \textrm{ for } \gf < \gf_0 
   \nonumber \\ \blanc
\auqe
Here we implicitly asumed that the profile $\gf(z)$ is a monotonous function of $z$ and accordingly
with $\gf_b^{(s, h)}$ = $\pi, 0$
the soft ($z < z_0$) and the hard ($z > z_0$) phases 
correspond to $\pi > \gf(z) > \gf_0$ and $\gf_0 > \gf(z) >0$ respectively. 
The value $\gf_0$ of $\gf$ at $z = z_0$ is obtained from the boundary condition (\ref{cond_lim_z0})
\cite{kronmuller_2002} 
\equa{cos_fi0}
  \cos (\gf_0) &=& \frac{1}{(1 - \ep_A\ep_K)} \left[ \fand h(1 - \ep_A\ep_J)  \right. \nonumber \\
              &+& \left. \left[ h^2(1 - \ep_A\ep_J)^2 + (1 - \ep_A\ep_K) \left[ (1 - \ep_A\ep_K) - 2h(1 + \ep_A\ep_J) \right]  \right]^{1/2}   
  \right]
\auqe
which is defined up to a critical value which defines the depinning field, {\it i.e.} the largest
value of $\gf_0$ which can be accomodated by the hard phase, and coincides with the coercive field 
$h_c$ since we expect the nucleation field of the soft phase to be smaller. $h_c$ is given by
\cite{kronmuller_2002}
\equa{h_c}
  h_c = \frac{1 - \ep_A\ep_K}{(1 + \sqrt{\ep_A\ep_J})^2} 
\auqe
%
%
When $z_b^{(s)}$ takes a finite value and then coincides with the half thickness of the soft layer,
$\gf_b = \gf_b(z_b^{(s)})$ in
equation (\ref{phi_1}) is considered as a parameter. The profile in equation (\ref{profil_1}) is
numerically integrated and $\gf_b$ is determined from the fulfillment of
\equa{z_fib}
z_b^{(s)} - z_0 &=& 
  - \int_{\gf_0}^{\gf_b} \sqrt{A_s}\left[ \gD U_{s}(\gf, \gf_b) \right]^{-1/2} d\gf ~~\equiv~I(\gf_b)
\nonumber \\ \textrm{ ~with~ } \cos(\gf_0) &=& 
 h(1 - \ep_A\ep_J) + \left[ h^2(\ep_A\ep_J - 1)^2 +1 - 2h(1 - \ep_A\ep_J\cos\gf_b)
                     \right]^{1/2}
\auqe
from a Newton-Raphson procedure by solving : $I(\gf_b)~-~(z_b^{(s)}~-~z_0)~=~0$.
%

Let us now consider the case where the easy axis of the hard inclusion is not oriented parallel to 
the interface, $(\hat{n},~\hat{z}) = \theta_h \neq \pi/2$, still with $(\hat{n},~\hat{y}) = \pi/2$. 
As a general rule, for ($\theta_h\;-\;\pi/2$) not too small,
the numerical simulations lead to a N\'eel domain wall at the interface, and accordingly we consider
only this situation in the one dimensional model. The magnetization is therefore in the $(\hat{x},~\hat{z})$
plane and is totally determined by the angle $\gf(z)$ = $(\hat{m},~\hat{z})$. As a consequence of 
the longitudinal nature of the domain wall at the soft / hard interface the demagnetizing field in the soft phase
must be introduced. We still do not introduce the demagnetizing field in the hard phase,
since we aim {\it in fine} to model a system where this latter concerns a cubic inclusion characterized by 
$N_x = N_y = N_z$. The soft phase is enclosed in an ellongated shaped prism, 
whose long axis, $\hat{z}$, is normal to the interface. In the following we shall consider 
the 1-D model in the infinitelly long prism geometry with demagnetizing coefficients 
$N_x = N_y = 1/2; N_z = 0$. In the bulk soft phase and if $z_b^{(s)}~\tend~\infty$, the equilibrium value $\gf(z_b^{(s)})$
is determined from the minimum of the energy density
\equa{ener_soft_1}
E_b = -J_sH_a \sin(\gf_b^{(s)}) + \frac{1}{2} \gD N \frac{J_s^2}{\mu_0} \sin^2(\gf_b^{(s)}); 
    \textrm {~~~with~~} \gD N = (N_x - N_z) = \frac{1}{2}
\auqe
with the result
\equa{phi_b}
\sin(\gf_b^{(s)}) = -\inf \left( 1, \frac{k h}{\ep_J} \right); \textrm{ ~~~ with ~:~} H_a = -H_K h
\auqe
In the bulk hard phase, the equilibrium value, $\gf_b^{(h)}$ of $\gf(z)$ is determined in a similar way
and can expanded in the vicinity of $\theta_h$ because of the high value of $H_K$ and we get
\equa{phi_b_h}
\gf_b^{(h)} = \theta_h - \frac{h\cos\theta_h}{1 + h \sin\theta_h}
\auqe
In the present geometry, $\gD U_{h,s}(\gf(z))$ are now given by
\equa{delta_U}
  \gD U_{h}(\gf(z)) &=& A_h\left( \frac{\pi}{\de_B} \right)^2 \left( \sin^2(\theta_h - \gf(z)) 
                       + 2 h\left( \sin(\gf(z)) - \sin(\gf_b^{(h)})  \right) \right)   \nonumber  \\
  \gD U_{s}(\gf(z)) &=& A_s\left( \frac{\pi}{\de_B} \right)^2 \frac{\ep_J^2}{k \ep_A}
                     \left( \sin^2(\gf(z))      + \frac{2kh}{\ep_J} \sin(\gf(z))
                          - \sin^2(\gf_b^{(s)}) - \frac{2kh}{\ep_J}\sin(\gf_b^{(s)})\right ) 
                                              \nonumber \\
\auqe
where we have introduced the hardness factor $k~=~2\mu_0H_K/J_h$.
In a similar way to what have been done for the parallel oriented interface we get the profile 
$\gf(z)$ from the solvation of

\equa{profil_s}
z^* - z^*_0 = \frac{1}{\pi} \frac{\left( \ep_A~k \right)^{1/2}} {\ep_J}
            \int_{\gf(z)}^{\gf_0} \frac{d\gf}
{\left( \sin^2(\gf) + (2 k h/\ep_J) \sin(\gf) - \sin^2(\gf_b^{(s)}) - (2 k h/\ep_J) \sin(\gf_b^{(s)}) \right)^{1/2}}
                                   \nonumber \\
\auqe
in the soft phase and 
\equa{profil_h}
z^* - z^*_0 &=& \frac{1}{\pi} 
            \int_{\gf_0}^{\gf(z)} \frac{d\gf}{\left( \sin^2(\gf - \theta) +2 h(\sin(\gf) - \sin(\theta)) \right)^{1/2} }
                                                                       \nonumber \\
\auqe
in the hard phase. The value $\gf_0$ of $\gf(z=z_0)$ is determined from the boundary condition and results 
from the numerical solvation of
\equa{phi_0_orient}
\frac{ \ep_A\ep_J^2}{k} \left[ \sin(\gf_0)^2 + 2h(k/\ep_J)\sin(\gf_0) - \sin(\gf_b^{(s)})^2 - 2h(k/\ep_J)\sin(\gf_b^{(s)})
                        \right] \nonumber \\
                      - \left[ \sin(\gf_0 - \theta_h)^2 - \sin(\gf_b^{(h)} - \theta_h)^2 +2h(\sin\gf_0 - \sin\gf_b^{(h)})
                        \right]
\nonumber \\
\auqe
As in the preceding case, when $z_b^{(s)}$ takes a constant value, the value of $\gf_b^{(s)}$ is no more taken 
from the equilibrium condition (\ref{phi_b}) but is determined from the fulfillment of $z(\gf_b^{(s)})~=~z_b^{(s)}$
with $z(\gf_b^{(s)})$ numerically calculated from equation (\ref{profil_s}).

Furthermore, we also introduce a way to fit the value of the field, say $h^{(fit)}$, at which the 1-D 
profile is calculated in order to enhance the accuracy with the 3-D simulated profile: instead of 
using the actual value of $h$, we fix the value of $\gf_b^{(s)}$ to that obtained in the simulation.
Hence, $h^{(fit)}$ satisfies 
\equa{fit}
\gf_b^{(s)}(h^{(fit)}) = \tilde{\gf}_b^{(s)}(h) ~~,
\auqe
$\tilde{\gf}_b$ being the 3-D micromagnetic simulation result for $\gf_b$.

The angular profile across the interface can be used to understand the behavior of the 
demagnetization process in exchange coupled media. This has been done in different situations 
in the literature \cite{fullerton_1998,dobin_2006,dobin_2007,pellicelli_2010}, 
especially in the framework of recording media optimisation although the magnetization 
profile accross the interface is generally not explicited. Here, our purpose is to make the 
link with the situation of a lattice of magnetically hard inclusions in a magnetically 
soft matrix. More precisely, we now compare the $\gf(z)$ profile calculated from equations 
(\ref{profil_2}), (\ref{profil_s}, \ref{profil_h}) with the one extracted from the 3-D 
micromagnetic simulation in two simple situations. 

\section {FEM simulation of the grain / matrix interface}
\label   {fem_3d}
 
The numerical simulation of the magnetization in terms of the applied field $H_a$ is
performed by the micromagnetic code MAGPAR \cite{magpar_2003}, 
based upon a finite elements numerical scheme. The embedded hard grain is a cube 
of edge length $a = 2R$, with $R$ = 17 $nm$ being the length scale fixed at a convenient 
value for nanostructured rare earth - transition metal intermetallics
\cite{chen_1999,chen_2000,hadjipanayis_1999,younsi_2010}. 
We have chosen to keep fixed the magnetic parameters of the hard phase and consequently
the Bloch domain wall thickness, $\de_B$ takes a constant value, $\de_B = 5 nm$
(see section \ref{resul} below). 
In the model including only one such embedded grain in order to represent the case
of isolated crystallites in the soft matrix this latter is a parallepipedic prism of 
total length $L_z$ = 6$R$, and lateral width $L_x = L_y = (2R + \de)$, with 
$\de = 0.40~\de_B$ and the embedded grain is located at its center.
In the model with two crystallites introduced in order to study the influence of 
$\gD/l_{ex}$ on the magnetization profile, we use a fixed edge to edge distance $\gD$
= 2$\de_B$ and the ratio $\gD/l_{ex}$ is varied through the value of $l_{ex}$ = 
$\sqrt{2\mu_0A_s/J_s^2}$ $\equiv~\de_B/\pi\sqrt{\mu_0 (H_K /J_h)(\ep_A/\ep_J^2)}$
as a function of $\ep_A$ and $\ep_J$.  
The local magnetization profile is extracted along a line parallel to the $\hat{z}$ 
direction, cutting the grain / matrix interface at its center. A particular 
attention is paid to the quality of the mesh, which must be superior to what is necessary 
for the average magnetization curve, $M(H_a)$ over the whole system. Here the quality of 
the mesh has been controlled through the fulfillment of the unitary condition of 
$\bf{m}(\bf{r})$ which is exactly satisfied only on the nodes. The typical size of the 
mesh tetraedra is sharply peaked at 0.10~$\de_B$, the maximum edge length for more than 
half of the tetraedra is less than 0.20~$\de_B$ leading to a mesh for our 
$(2R + \de)\times(2R + \de)\times 6R$ model including about 8.5$\;10^5$ finite elements.

Before going further we have to note that because in our system
the soft phase surrounds the whole hard grain, we have to take into account 
not only the interface we are interested in, normal to $\hat{z}$, but also the 
other sides of the embedded cubic grain. By symmetry, only the top and bottom ones are 
to be considered. This is important especially for the nucleation field determination. 
To estimate the effect of the additional interface, we divide the soft phase
domain into the prism based on the side normal to $\hat{z}$ located at $z = z_0$
namely bounded by $(x, y) = \pm R$, and the top and bottom layers ($\abs{x} > R$) 
for which the hard/soft interface is perpendicularly oriented.
For these additional layers, the demagetizing factors are $N_y = N_z \simeq 0$ and $N_x \simeq 1$
and we must add to the energy density a term due to the corresponding demagnetizing field and
proportional to the volumic fraction say $\al$ occupied by these layers in the total soft phase domain. 
Then, in the case of the Bloch wall parallel to the interface, the implicit 
equation from which $h_{nucl}$ is obtained \cite{skomski_1993,asti_2006,ghidini_2007},
\equa{h_nucl}
\gl_h\tanh(\gl_h L_h/2) = \ep_A\gl_s\tan(\gl_s L_s/2)
\auqe
still holds but with modified definitions of the parameters $\gl_h$ and $\gl_s$
\equa{la_sh}
\gl_h^2 &=&   \left( \frac{\pi}{\de_B} \right)^2 \left[ 1 - h  \right]
\nonumber \\
\gl_s^2 &=&  \left( \frac{\pi}{\de_B} \right)^2 \left( \frac{\ep_J }{\ep_A} \right) 
             \left[ h + \al \frac{N_x J_s}{\mu_0 H_K} - \ep_K/\ep_J  \right]
\auqe
Since $\gl_h L_h > 1$ for $L_h > \de_B$ the value of $h_{nucl}$ is nearly independent of $L_h$, 
and therefore, we can estimate 
the effect of the surrounding layers by solving equation (\ref{h_nucl}) for the field $h' \simeq h + \al N_z J_s/\mu_0 H_K$.
The nucleation field is thus approximately given by $h_{nucl} = h_0 - \al N_z J_s/\mu_0 H_K$ where $h_0$ is the solution
of equ. (\ref{h_nucl}) with $\al = 0$, and may be negative corresponding to a nucleation of the soft phase
in the first quadrant. For the same geometrical reason, we do not expect the coercive field to be given
by the analytical result of the 1-D model (see equ. (\ref{h_c})).

In the following, we mainly focus on the local magnetization profile in term of $z$ for different 
values of the reduced external field, $h$. 

\section {Results and discussion}
\label   {resul}
\begin{figure}[h] 
\includegraphics*[ angle = -00, width = 0.70\textwidth]{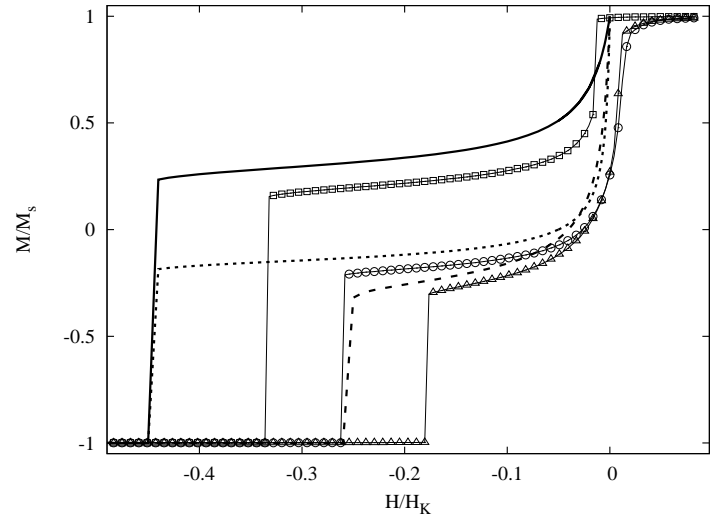} 
\caption{ \label{hyst_1c} 
Demagnetization curve as calculated by using the profile calculated on the 1-D model
(lines) and compared to the results of the 3-D micromagnetic simulations
(symbols).
($\ep_A, \ep_J$) = (0.75, 0.325) open squares and solid line; 
(0.325, 0.75) open circles and dotted line; 
(1.00, 1.00) triangles and dashed line.
} 
\end{figure}

\begin{figure}[h] 
\blanc \vskip -0.05\textheight
\includegraphics[ angle = -00, width = 0.70\textwidth]{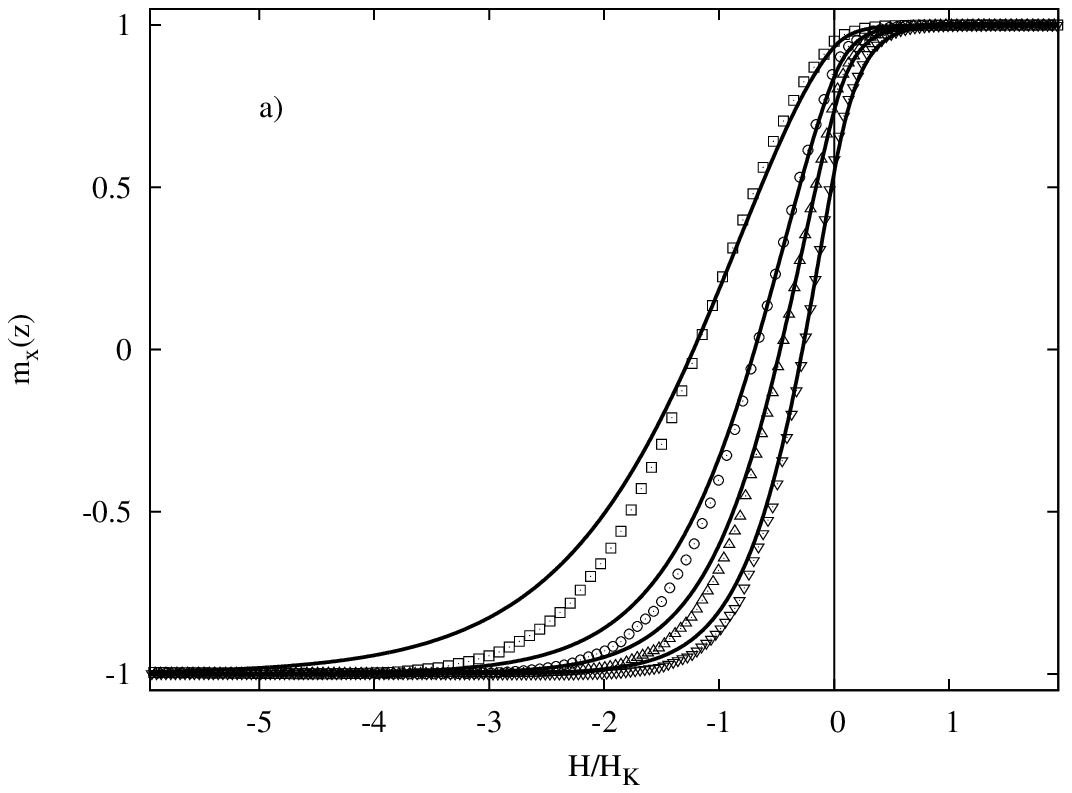} 
\\ \vskip -0.01\textheight
\includegraphics[ angle = -00, width = 0.70\textwidth]{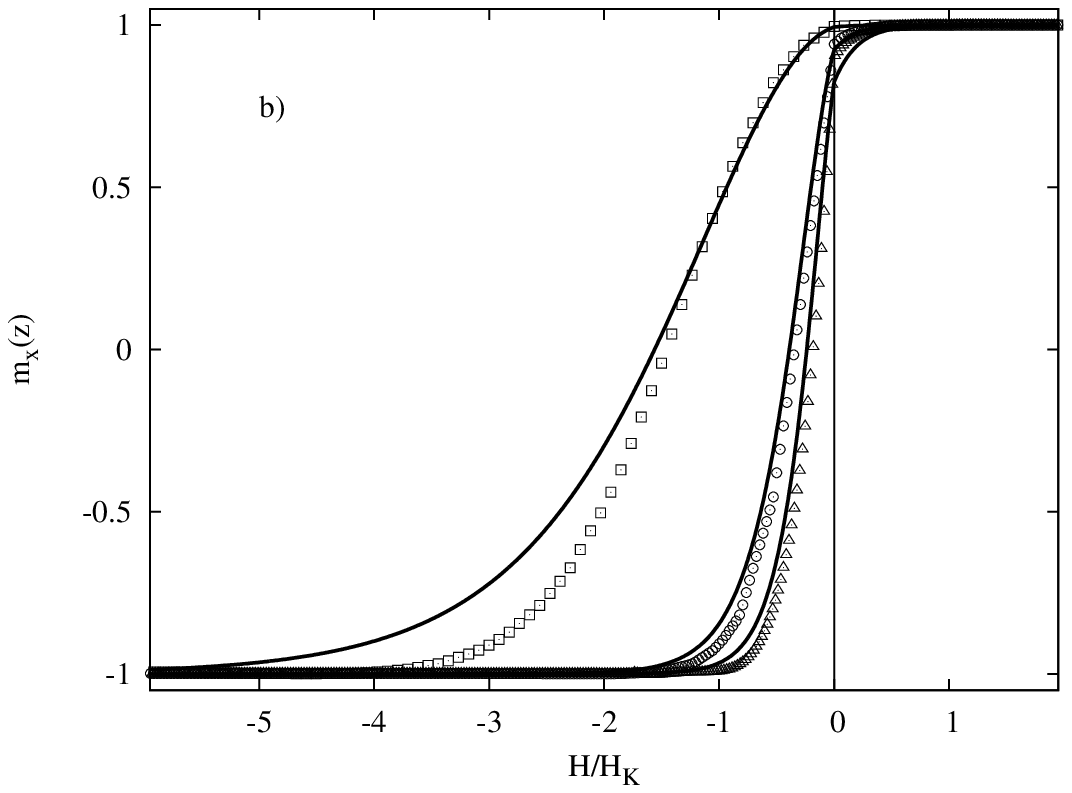} 
\\ \vskip -0.01\textheight
\includegraphics[ angle = -00, width = 0.70\textwidth]{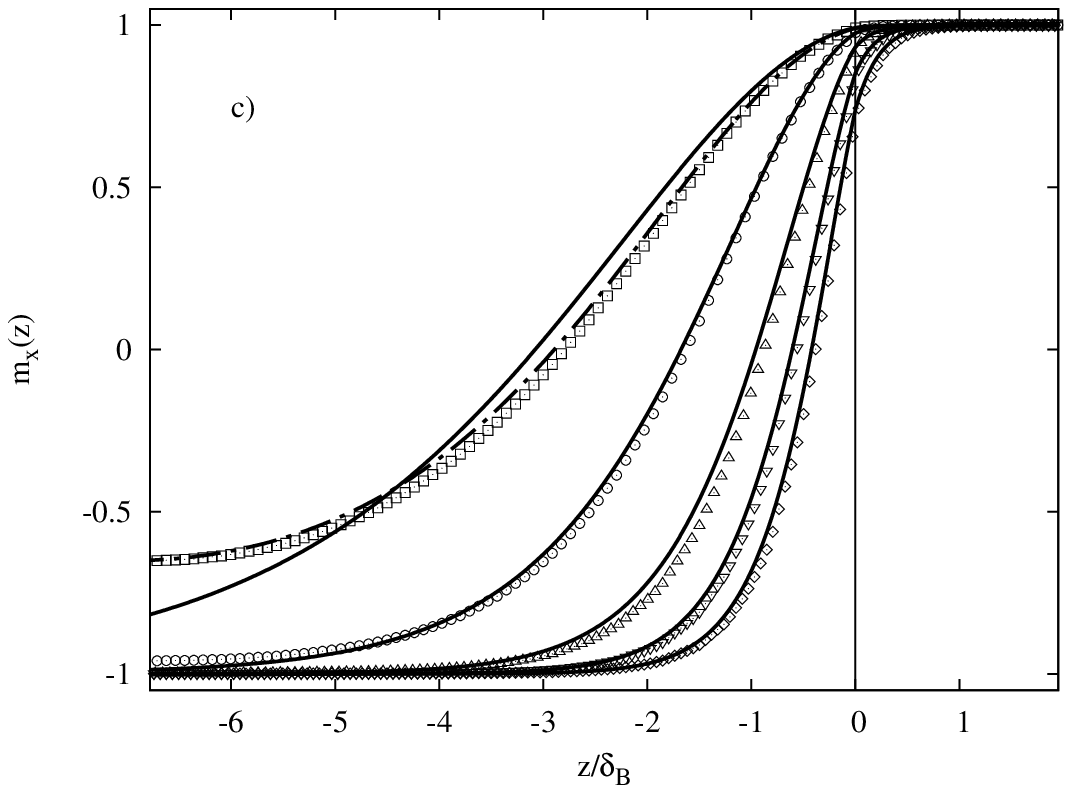} 
\caption{ \label{fig_prof_1c} 
Magnetization profile, $m_x(z)$ in terms of $z/\de_B$ across the hard/soft interface along 
the normal cutting the center of the cubic inclusion. The hard (soft) phase is located
at $z > 0$ ($z < 0$).
Solid line: result of the 1-D model; Symbols: 3-D micromagnetic simulation.
$K_s = 0; K_h = 3.05~10^6~Jm^{-3}$~; \\ 
a)~:~$\ep_A = 1,~\ep_J = 1$         ~$h$ = 0.0328, squares; 0.0740, circles; 0.1150, upward triangles; 0.1760, downward triangles.
b)~:~$\ep_A = 0.325,~\ep_J = 0.75$  ~$h$ = 0.0123, squares; 0.1352, circles; 0.2580, triangles.
c)~:~$\ep_A = 0.75,~\ep_J  = 0.325$ ~$h$ = 0.0165, squares; 0.0492, circles; 0.1230, upward triangles; 0.2295, downward triangles; 0.3320, diamonds.
Dash-dotted line : result of the 1-D model calculated with $z_b^{(s)}/\de_B$ = 7.0 
and a fitted value of the field $h^{(fit)}$ = 0.0240, according to equation (\ref{fit}).
} 
\end{figure}

We first focus on the magnetization profile obtained at the interface between
a cubic grain of magnetically hard material and the soft matrix. We consider a magnetically
hard material characterized by $K_h = 3.05\;10^{6}Jm^{-3}$, $A_h = 7.7\;10^{-12}J/m^{-1}$
and $J_h = 1T$, typical values for the Re-Fe compounds \cite{hadjipanayis_1999},
leading to $\de_B$ = 5$nm$, the soft matrix being characterized by $K_s$ = 0 and 
the other parameters defined through the values of $\ep_A$ and $\ep_J$.
We first compare the demagnetization curves given by the true 3-D model to that one should get from
the 1-D profile as calculated from equation (\ref{profil_2}). The latter is obtained analytically from 
the magnetization profile and the geometrical average weighted by the respective volumes of the hard
and soft phases in the 3D model. The result is displayed in figure (\ref{hyst_1c}) for 3 
characteristic sets of values of ($\ep_A, \ep_J$). As can be seen, the agreement is qualitatively 
correct, the main discrepancy being the values of the nucleation and coercive fields. This stems 
from shape and finite size effect as as been discussed for the nucleation field above. On the other 
hand, the characteristic plateau in between $h_{nucl}$ and $h_c$ is very well reproduced 
especially concerning its $h-$ dependence. Then, in figure (\ref{fig_prof_1c}) we compare the 
magnetization profile $m_z(z) = \cos(\gf(z))$ as calculated in the 1-D description corresponding 
to $z_b^{(s)}~=~\infty$ to the one obtained from the 3-D simulation. This is done for a set of 
values of $h$ corresponding to characteristic points on the demagnetization plateau 
($h_{nucl} < h < h_c$). We see that on the one hand the compression process of the domain wall in
the soft side of the interface is clearly evidenced and on the other hand the agreement between
the 3-D simulated profile and the 1-D anaytical one is quite satisfactory. The only case where 
the calculated 1-D magnetization profile deviates from the simulation result at large values of
$\abs{z}$ corresponds to a situation ($\ep_A = 0.75; \ep_J = 0.325$ and $l_{ex}$ = 2.345~$\de_B$) 
where the plateau $m_x(z)$ = -1 is not reached at the boundary of the system (see figure 
(\ref{fig_prof_1c}c)) because the value of the field is too small to compress
the profile in the limits of the micromagnetic model. We note that the boundary of the system 
corresponds to $\abs{z/\de_B}~=~6.8$. As can be seen in figure (\ref{fig_prof_1c}) the fitting 
procedure introduced through equation (\ref{fit}) definitely solves the problem. 
We conclude that the 1-D model leads to a rather good 
approximation for the magnetisation curve, and a very good approximation for the angular profile at 
the hard/soft interface. Accordingly the local demagnetization process at the hard grain / soft
matrix interface follows the one deduced from the 1-D layered model. In particular, we emphasize that
the strong reduction of the coercive field due to the exchange coupling is reproduced
although the value of $h_c$ differs due to the mentioned shape and size effects resulting from
the embedding geometry.

\begin{figure}[h] 
\includegraphics[ angle = -00, width = 0.70\textwidth]{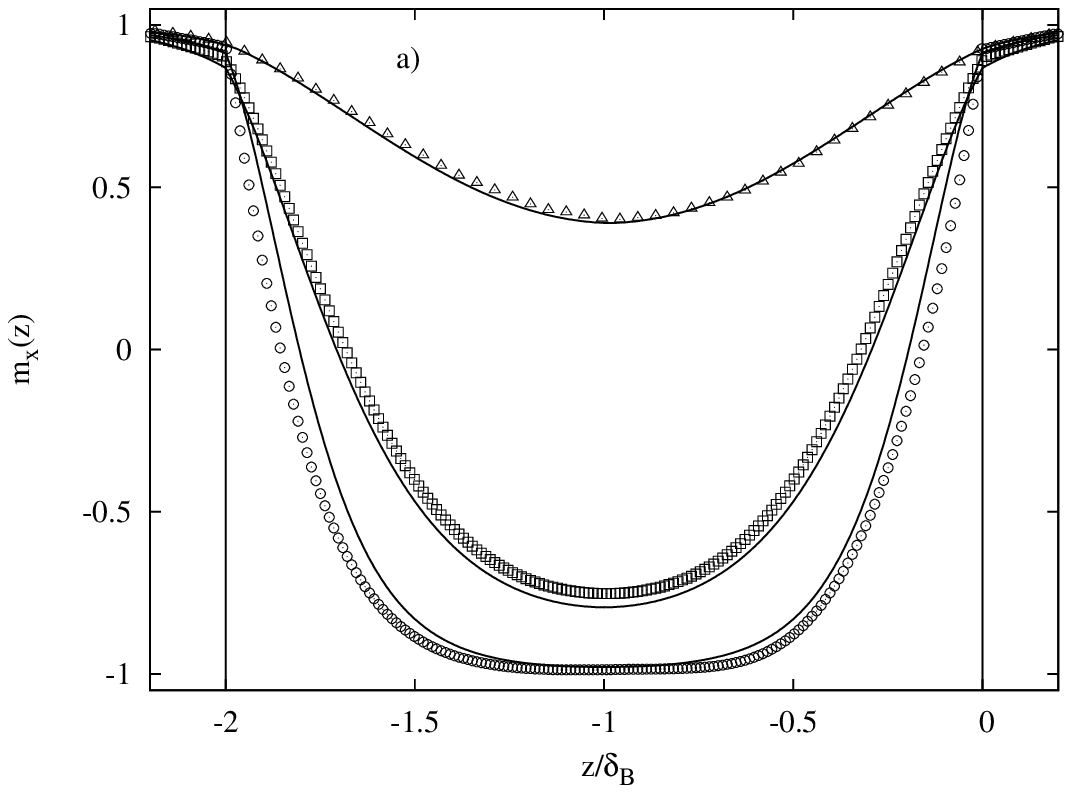} 
\\
\includegraphics[ angle = -00, width = 0.70\textwidth]{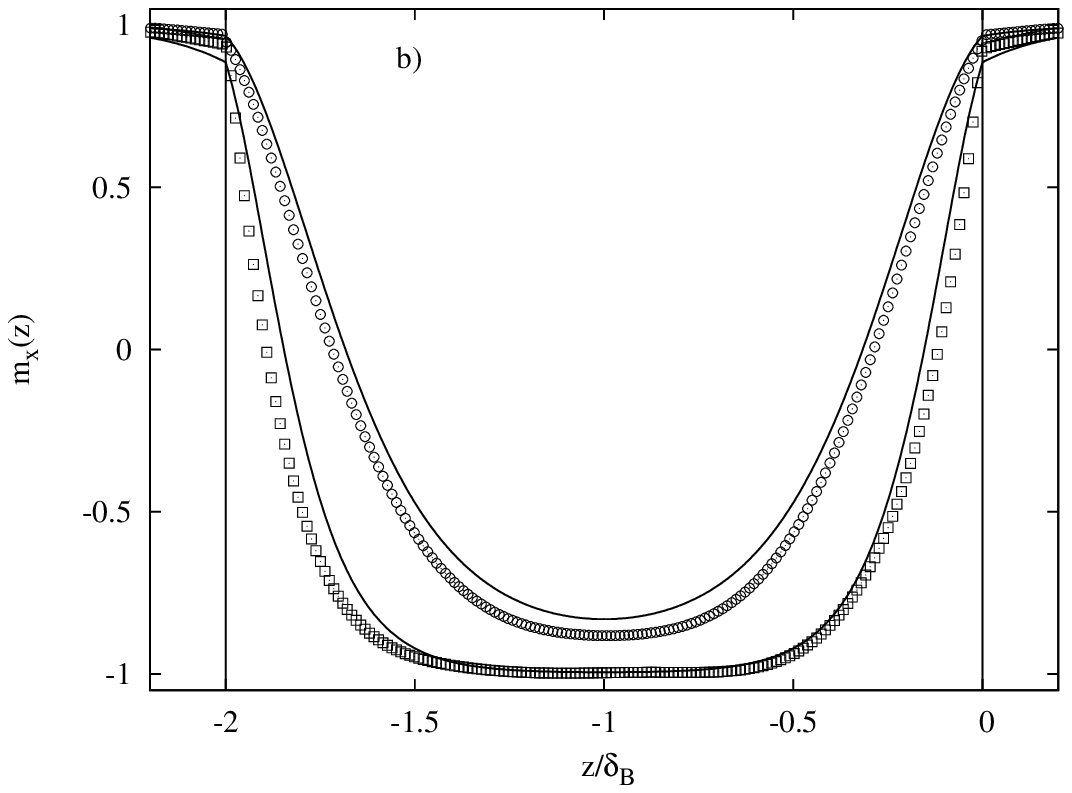} 
\caption{ \label{fig_prof_2c} 
Magnetization profile,$m_x(z)$ between the two cubic inclusions in the soft matrix.
$\gD$ = 2$\de_B$.
The center of soft layer is located at $z/\de_B = -1.0$. Symbols: 3-D micromagnetic simulation;
Solid line: profile as calculated from the 1-D model.
\\ a)~ $\ep_J$ = 0.75 and :
$\ep_A$ = 0.75 and $h$ = 0.1475, triangles; $\ep_A$ = 0.325 and $h$ = 0.2295, squares; $\ep_A$ = 0.162 and $h$ = 0.2622, circles. 
For $\ep_A$ = 0.75, the 1-D model profile is calculated with the fitted value of the field, 
$h^{(fit)}$ = 0.1874, according to equation (\ref{fit}).
In each case, the value of the field is close to the depinning field of the 3-D micromagnetic simulation.
\\ b)~ $\ep_J$ = 1.50~; $\ep_A$ = 0.1623 and $h$ = 0.0697, circes; 0.1885, squares. 
The larger value of the field is close to the depinning field of the 3-D micromagnetic simulation.
}
\end{figure}

\begin{figure}[h] 
 \includegraphics[ angle = -00, width = 0.70\textwidth]{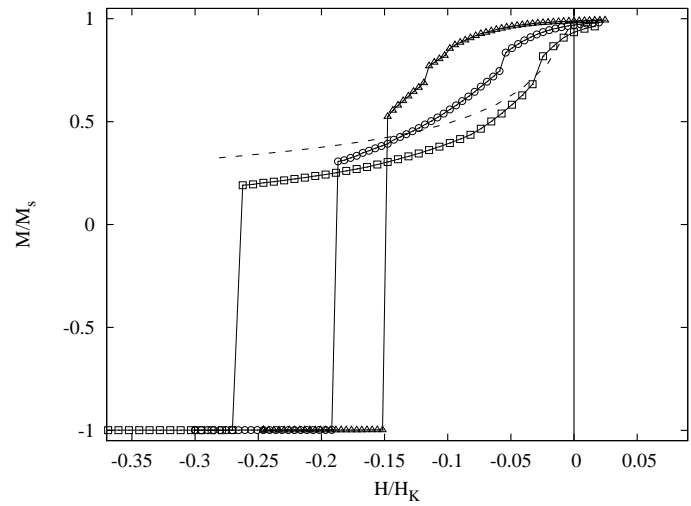}
 \caption{ \label{hyst_2_cube} 
Demagnetization curve of the two cubic inclusions model.
$\ep_J$ = 0.75~; $\ep_A$ = 0.162, open squares; 0.325, open circles; 0.75 open triangles.
Dashed line~: demagnetization curve of the 1-D model for the hard/soft interface with
$\ep_J$ = 0.75 and $\ep_A$ = 0.162.
  }
 \end{figure}

Now we focus on the influence of the edge to edge distance $\gD$ between 
two inclusions, still in the parallel orientation, namely $\hat{n} = \hat{x}$ for the two  
inclusions. The 1-D angular profile is calculated by using the numerical determination of 
$\gf_b^{(s)}$ as described in section \ref{1d_model}, equation (\ref{z_fib}). 
The magnetization profile $m_x(z)$ for the value of the field closest
to the depinning point of the 3-D simulation is displayed in figure (\ref{fig_prof_2c}). 
As $\ep_A$ increases from 
$\ep_A$ = 0.162 to 0.75, the ratio $(\gD/2)/l_{ex}$ of the distance between the mid-plane 
and the interface to the exchange length decreases from 2.115 to 0.983; hence 
the flexibility of the profile $m_x(z)$ in the soft phase is sufficient for this latter
to reach a plateau at $m_x(z)$ = -1 in between the two hard inclusions only in the first case.
Moreover when $(\gD/2)/l_{ex} \leq 1.0$ the two inclusions become exchange coupled {\it via}
the soft phase, and the magnetization reversal occurs as a whole with a one phase like
behavior, as can be seen in figure (\ref{hyst_2_cube}) where the corresponding demagnetization 
curve is displayed. This means that in this case, a strong coupling regime is reached.
This is in qualitative agreement with the magnetic phase diagram of Ref. \cite{asti_2004}
since we clearly evidence the three phases namely the decoupled magnet for
$(\gD/2)/l_{ex} > 2$, the ES coupled magnet for $2 > (\gD/2)/l_{ex} > 1$ and the rigid
magnet for $(\gD/2)/l_{ex} < 1$. 

\begin{figure}[h] 
 \includegraphics[ angle = -00, width = 0.70\textwidth]{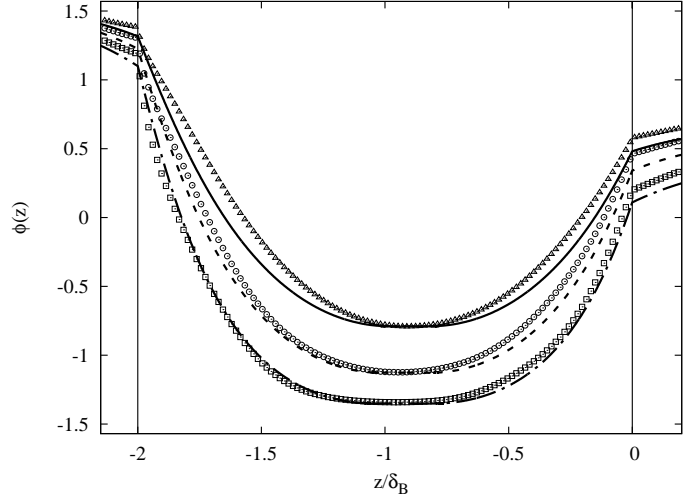}
 \caption{ \label{prof_pi_4_theta} 
Angular profile accross the soft phase between the two hard cubic inclusions located
at $z/\de_B~<~-2$ and $z/\de_B~>~0$ respectively with $(\hat{n},~\hat{z})~=~\pi/4$ for the latter.
Symbols: 3-D micromagnetic simulations for $h$~= 0.0655, triangles; 0.1311, circles 
and 0.3440, squares.
1-D profile for the fitted values of the field 
$h^{(fit)}$~= 0.130, solid lines; 0.2020, dashed lines and 0.3080, dash-dotted lines respectivelly. 
$h^{(fit)}$ is determined using equation (\ref{fit}) to fit  $\gf(z = z_b^{(s)})$ on the simulated result
where $z_b^{(s)}/\de_B~=~-1$ is the location of the soft layer mid plane.
 }
 \end{figure}

\begin{figure}[h] 
 \includegraphics[ angle = -00, width = 0.70\textwidth]{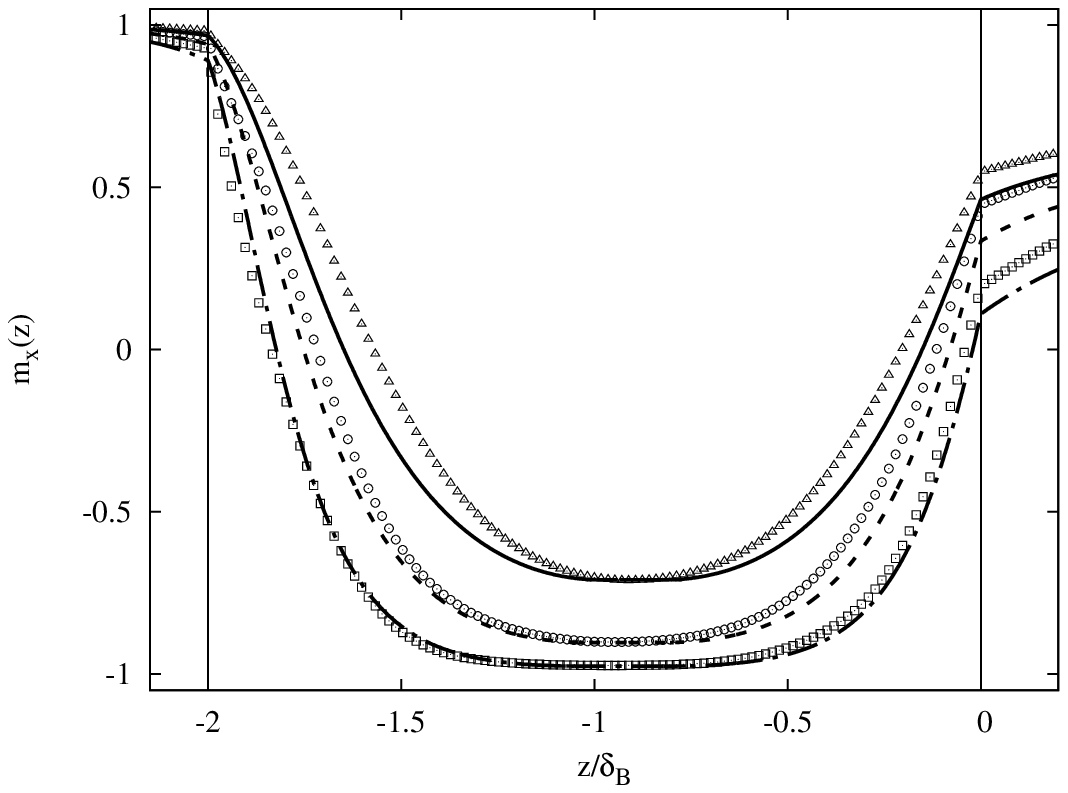}
 \caption{ \label{prof_pi_4_mx} 
Same as figure (\ref{prof_pi_4_theta}) for the $x-$~component of the magnetization $\hat{m}(z)$.
 }
 \end{figure}

\begin{figure}[h] 
 \includegraphics[ angle = -00, width = 0.70\textwidth]{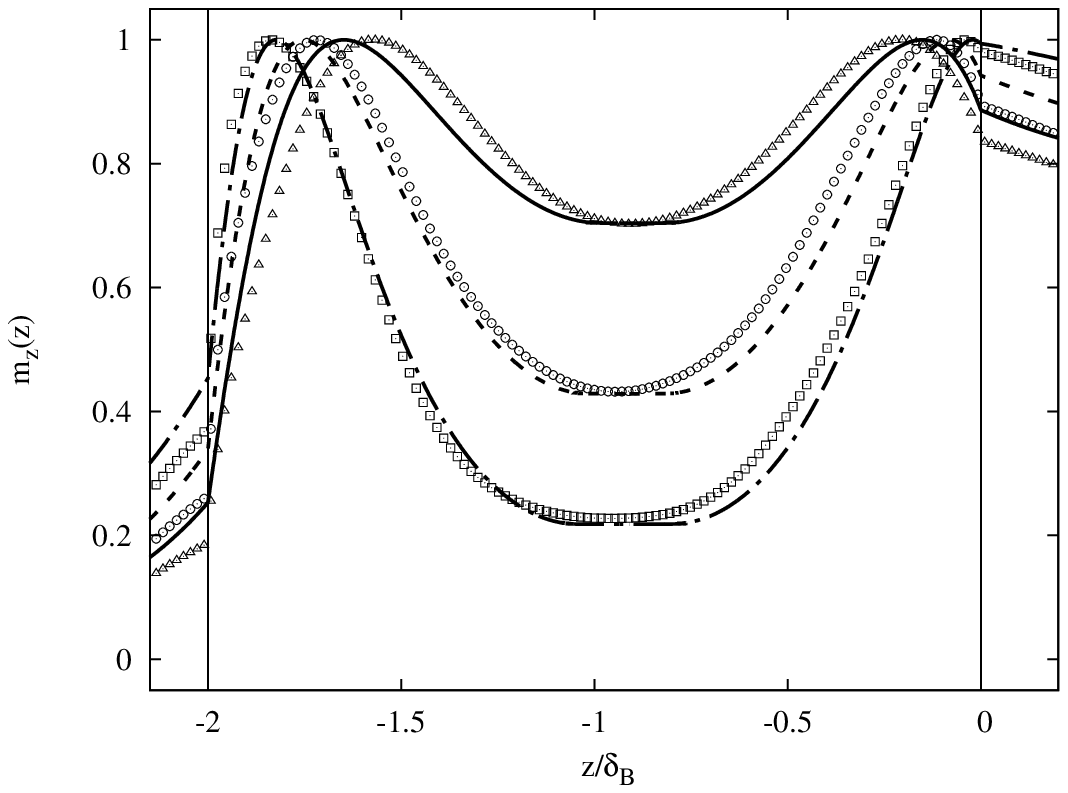}
 \caption{ \label{prof_pi_4_mz} 
Same as figure (\ref{prof_pi_4_theta}) for the $z-$~component of the magnetization $\hat{m}(z)$.
  }
 \end{figure}

The same conclusions as above hold for the case when one easy axis is not parallel to the
interface; Here we have chosen as an exemple $\theta_h = \pi/4$; the 3-D simulation has
been performed only in the two inclusions model with $\theta_h \ne 0$ for one of the 
inclusions. The results for the angular profile $\gf(z)$ and the magnetization profiles 
$m_x(z)$ and $m_z(z)$ are given in figures (\ref{prof_pi_4_theta}), (\ref{prof_pi_4_mx}) 
and (\ref{prof_pi_4_mz}) respectively. 
One important difference with the case $\theta_h = 0$ is that the magnetization of the
hard phase in the $\theta_h \ne 0$ grain presents a reversible variation of $m_x$
before switching as is the case in Stoner Wolfarth spherical particles when the easy
axis does not coincide with the external field direction.
The other difference as already mentioned is the fact that the hard~/~soft interface 
for $h < h_c$ is a longitudinal (N\'eel) domain wall instead of transverse (Bloch) one
as it is evidenced from the figure (\ref{prof_pi_4_mz}).

\begin{figure}[h] 
 \includegraphics[ angle = -00, width = 0.70\textwidth]{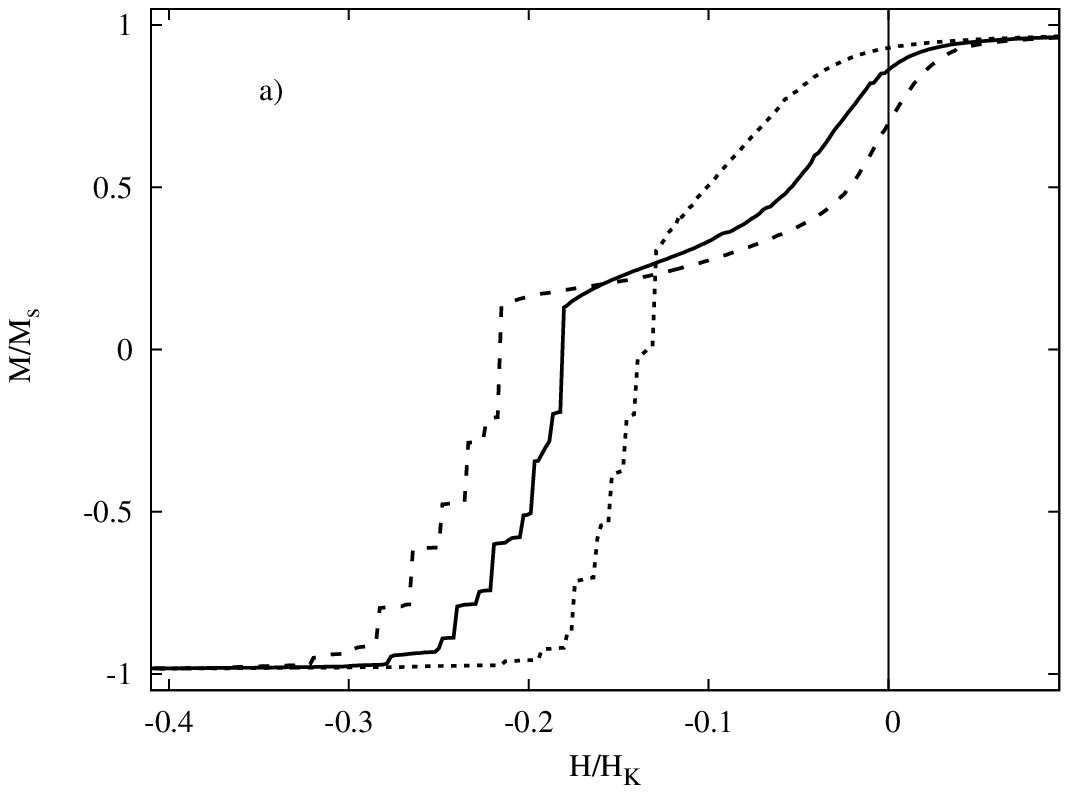}
 \\
 \includegraphics[ angle = -00, width = 0.70\textwidth]{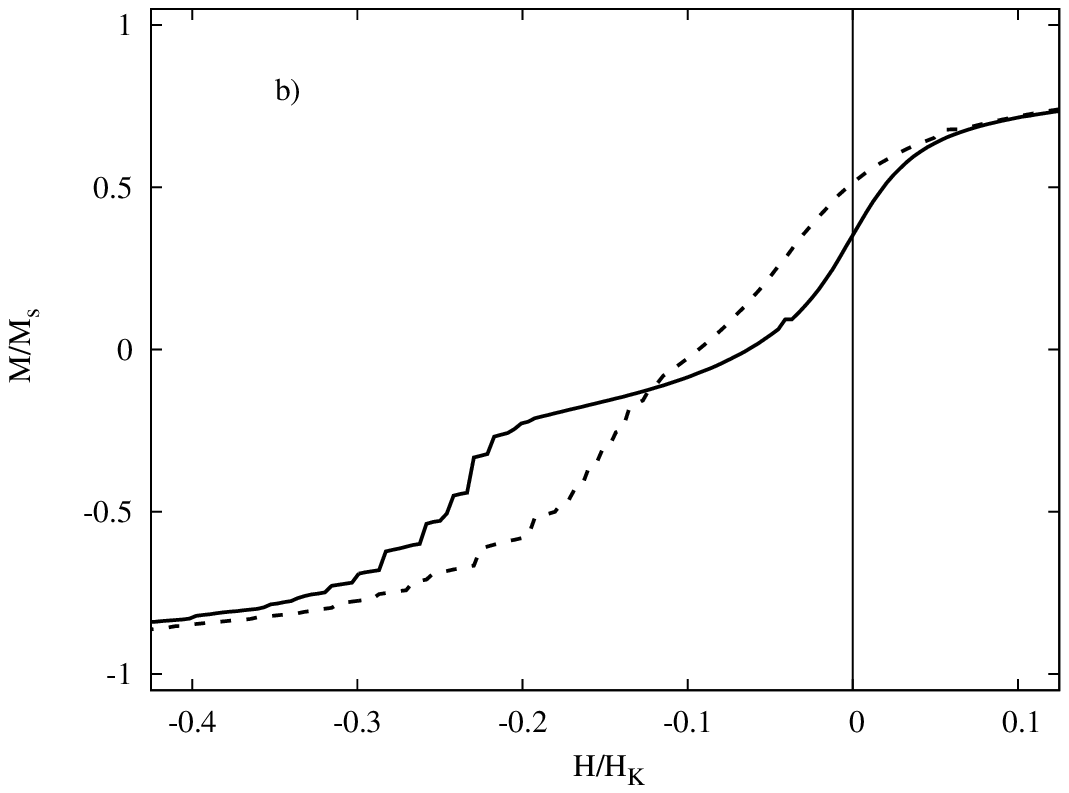}
 \caption{ \label{hyst_256} 
Demagnetization curve obtained from a micromagnetic simulation on the system including 
256 cubic particles located on the nodes of a simple cubic lattice.\\
a)~ Preferentially oriented easy axes in the direction of the 
field with $\Sigma_i(\abs{(\hat{n}_i.\hat{x})})/N_p = 0.94$~;
$\ep_J$ = 0.75 and $\ep_A$ = 0.162, dashed line; 0.325, solid line and 0.75 dotted line. \\
b)~ Random distribution of easy axes; $\ep_J$ = 0.75 and $\ep_A$ = 0.162, solid line; 0.75, dashed line.
  }
 \end{figure}

Now as we have shown from the local magnetization profile across the interface 
that the process for the demagnetization at the layered system holds
at the hard~/~soft interface of 3D hard inclusions in a soft matrix at least for
one or two finite sized objects, we relate the demagnetization curve of the
two inclusions model to an extended one. 
We consider a lattice including $N_p = 256$ cubic inclusions 
located on the nodes of a simple cubic lattice made of 4 (8 $\times$ 8) planes 
and is an extended version of the preceding two grains model. 
The edge to edge distance between nearest neighbors in each plane is 
$\gD = 2\de_B$ as above and the external field is along $\hat{x}$.
Two different easy axes distributions have been used; 
in the first one all the easy axes are close to the $\hat{x}$ -axis, 
with $\Sigma_i(\abs{(\hat{n}_i.\hat{x})})/N_p = 0.94$ and in the second one
the axes are randomly distributed on the unit sphere 
($\Sigma_i(\abs{(\hat{n}_i.\hat{x})})/N_p = 0.5$).
We have considered the set of parameters as in fig. (\ref{hyst_2_cube}).
Since 2/3$^{-rd}$ of the interfaces are parallel to both the preferential orientation
of the $\{\hat{n}_i\}$ and the direction of the applied field, the comparison
with the preceding model for the parallel interface is meaningful.
As we can see on figure (\ref{hyst_256}\;a), at the qualitative level, the demagnetization 
curve for the easy axes preferentially distributed along $\hat{x}$ is close the that
obtained for the 2- inclusions 'toy' model. The demagnetization curve calculated with the 
randomly distributed easy axes is displayed on figure (\ref{hyst_256}\;b), 
and leads to the same type of conclusion but only at a qualitative level.
Given the results we have got on the two inclusions 'toy' model, we conclude that the
caracteristic plateau in the demagnetization curve is indeed the signature of the
local magnetization profile compression / depinning process. However, considering a 
mesh of the required quality for extracting the magnetization profile in the
256 inclusions model would lead to prohibitively heavy simulations.
Indeed, our analysis of the local magnetization profile is based upon a continuous function
for the later and its behavior with respect to the external field. Such a requirement
from the finite element simulation results, where the profile is obtained from an
interpolation between the mesh nodes, can be reached only with a very fine mesh. 
Nevertheless, on the physical point of view, in spite of its continuous nature the 
magnetization profile resulting either from an analytical calculation or a finite 
elements interpolation cannot be interpreted at an infinitely small length scale 
since the micromagnetic formalism remains a continuous medium type of approach valid 
beyond some characteristic lenght scale, say $\sim$ 1 nm.

We conclude that as is the case on the small local model, the rigid magnet, exchange spring
and decoupled magnet regimes are reached on the extended model for roughly $\gD/(2l_{ex})~<~1$, 
$1~<\gD/(2l_{ex})~<~2$ and $\gD/(2l_{ex})~\geq~2$ respectively. In this case, these boundary 
values can be related to the volumic fraction $\gf_v$ of the embedded cristallites through 
$\gD/(2l_{ex})~=~(\gf_v^{-1/3} - 1)(R/l_{ex})$ with the simple cubic geometry.
This extended model provides a link with a more realistic modelling of an actual experimental
system as was shown in ref. \cite{younsi_2010}. Thus, the results displayed in figure
(\ref{hyst_256}) show that the local demagnetization process explicited at the single
inter-grains soft layer can be qualitatively transfered to a realistic situation.
\section {Conclusion}

In this work, focusing on the well established mecanism of exchange coupling between phases
in hard~/~soft composite systems, we made the connection between the demagnetization
behavior at the layered system where a 1-D description holds and a fully 3-D composite system.
The connection has been done through the local magnetization profile and we have shown that
locally at the hard~/~soft interface this profile can be transfered from the layered to the
3-D systems. 
While it is well known that the distance between hard objects in a soft matrix as measured 
w.r.t. the soft phase exchange length is one of the most relevant parameters driving the coupling,
this point on the one hand has been made more quantitative and on the hand illustrated by 
the behavior of the simulated local magnetization profile in a small 3-D model. 
This latter play the role of an intermediate between the 1-D description of the interface and 
the true 3-D model.

\section*{Acknowledgements}
The numerical micromagnetic simulations were performed using HPC resources from GENSI-CINES
(grant number 2011-096180).
%

\begin{thebibliography}{26}
\expandafter\ifx\csname natexlab\endcsname\relax\def\natexlab#1{#1}\fi
\expandafter\ifx\csname bibnamefont\endcsname\relax
  \def\bibnamefont#1{#1}\fi
\expandafter\ifx\csname bibfnamefont\endcsname\relax
  \def\bibfnamefont#1{#1}\fi
\expandafter\ifx\csname citenamefont\endcsname\relax
  \def\citenamefont#1{#1}\fi
\expandafter\ifx\csname url\endcsname\relax
  \def\url#1{\texttt{#1}}\fi
\expandafter\ifx\csname urlprefix\endcsname\relax\def\urlprefix{URL }\fi
\providecommand{\bibinfo}[2]{#2}
\providecommand{\eprint}[2][]{\url{#2}}

\bibitem[{\citenamefont{Chen et~al.}(1999)\citenamefont{Chen, Meng-Burany, and
  Hadjipanayis}}]{chen_1999}
\bibinfo{author}{\bibfnamefont{Z.}~\bibnamefont{Chen}},
  \bibinfo{author}{\bibfnamefont{X.}~\bibnamefont{Meng-Burany}},
  \bibnamefont{and} \bibinfo{author}{\bibfnamefont{G.~C.}
  \bibnamefont{Hadjipanayis}}, \bibinfo{journal}{Applied Physics Letters}
  \textbf{\bibinfo{volume}{75}}, \bibinfo{pages}{3165} (\bibinfo{year}{1999}).

\bibitem[{\citenamefont{Chen et~al.}(2000)\citenamefont{Chen, Meng-Burany,
  Okumura, and Hadjipanayis}}]{chen_2000}
\bibinfo{author}{\bibfnamefont{Z.}~\bibnamefont{Chen}},
  \bibinfo{author}{\bibfnamefont{X.}~\bibnamefont{Meng-Burany}},
  \bibinfo{author}{\bibfnamefont{H.}~\bibnamefont{Okumura}}, \bibnamefont{and}
  \bibinfo{author}{\bibfnamefont{G.~C.} \bibnamefont{Hadjipanayis}},
  \bibinfo{journal}{Journal of Applied Physics} \textbf{\bibinfo{volume}{87}},
  \bibinfo{pages}{3409} (\bibinfo{year}{2000}).

\bibitem[{\citenamefont{Younsi et~al.}(2010)\citenamefont{Younsi, Russier, and
  Bessais}}]{younsi_2010}
\bibinfo{author}{\bibfnamefont{K.}~\bibnamefont{Younsi}},
  \bibinfo{author}{\bibfnamefont{V.}~\bibnamefont{Russier}}, \bibnamefont{and}
  \bibinfo{author}{\bibfnamefont{L.}~\bibnamefont{Bessais}},
  \bibinfo{journal}{Journal of Applied Physics} \textbf{\bibinfo{volume}{107}},
  \bibinfo{eid}{083916} (\bibinfo{year}{2010}).

\bibitem[{\citenamefont{Hadjipanayis}(1999)}]{hadjipanayis_1999}
\bibinfo{author}{\bibfnamefont{G.~C.} \bibnamefont{Hadjipanayis}},
  \bibinfo{journal}{Journal of Magnetism and Magnetic Materials}
  \textbf{\bibinfo{volume}{200}}, \bibinfo{pages}{373 } (\bibinfo{year}{1999}).

\bibitem[{\citenamefont{Kneller and Hawig}(1991)}]{kneller_1991}
\bibinfo{author}{\bibfnamefont{E.}~\bibnamefont{Kneller}} \bibnamefont{and}
  \bibinfo{author}{\bibfnamefont{R.}~\bibnamefont{Hawig}},
  \bibinfo{journal}{IEEE Trans. Magn.} \textbf{\bibinfo{volume}{27}},
  \bibinfo{pages}{3588} (\bibinfo{year}{1991}).

\bibitem[{\citenamefont{Fullerton et~al.}(1998)\citenamefont{Fullerton, Jiang,
  Grimsditch, Sowers, and Bader}}]{fullerton_1998}
\bibinfo{author}{\bibfnamefont{E.~E.} \bibnamefont{Fullerton}},
  \bibinfo{author}{\bibfnamefont{J.~S.} \bibnamefont{Jiang}},
  \bibinfo{author}{\bibfnamefont{M.}~\bibnamefont{Grimsditch}},
  \bibinfo{author}{\bibfnamefont{C.~H.} \bibnamefont{Sowers}},
  \bibnamefont{and} \bibinfo{author}{\bibfnamefont{S.~D.} \bibnamefont{Bader}},
  \bibinfo{journal}{Phys. Rev. B} \textbf{\bibinfo{volume}{58}},
  \bibinfo{pages}{12193} (\bibinfo{year}{1998}).

\bibitem[{\citenamefont{Victora and Shen}(2005)}]{victora_2005}
\bibinfo{author}{\bibfnamefont{R.}~\bibnamefont{Victora}} \bibnamefont{and}
  \bibinfo{author}{\bibfnamefont{X.}~\bibnamefont{Shen}},
  \bibinfo{journal}{IEEE Trans. Magn.} \textbf{\bibinfo{volume}{41}},
  \bibinfo{pages}{2828} (\bibinfo{year}{2005}).

\bibitem[{\citenamefont{Skomski et~al.}(2008)\citenamefont{Skomski, George, and
  Sellmyer}}]{skomski_2008}
\bibinfo{author}{\bibfnamefont{R.}~\bibnamefont{Skomski}},
  \bibinfo{author}{\bibfnamefont{T.~A.} \bibnamefont{George}},
  \bibnamefont{and} \bibinfo{author}{\bibfnamefont{D.~J.}
  \bibnamefont{Sellmyer}}, \bibinfo{journal}{Journal of Applied Physics}
  \textbf{\bibinfo{volume}{103}}, \bibinfo{eid}{07F531} (\bibinfo{year}{2008}).

\bibitem[{\citenamefont{Suess}(2006)}]{suess_2006}
\bibinfo{author}{\bibfnamefont{D.}~\bibnamefont{Suess}},
  \bibinfo{journal}{Applied Physics Letters} \textbf{\bibinfo{volume}{89}},
  \bibinfo{eid}{113105} (\bibinfo{year}{2006}).

\bibitem[{\citenamefont{Suess}(2007)}]{suess_2007}
\bibinfo{author}{\bibfnamefont{D.}~\bibnamefont{Suess}},
  \bibinfo{journal}{Journal of Magnetism and Magnetic Materials}
  \textbf{\bibinfo{volume}{308}}, \bibinfo{pages}{183 } (\bibinfo{year}{2007}).

\bibitem[{\citenamefont{Goncharov et~al.}(2007)\citenamefont{Goncharov,
  Schrefl, Hrkac, Dean, Bance, Suess, Ertl, Dorfbauer, and
  Fidler}}]{goncharov_2007}
\bibinfo{author}{\bibfnamefont{A.}~\bibnamefont{Goncharov}},
  \bibinfo{author}{\bibfnamefont{T.}~\bibnamefont{Schrefl}},
  \bibinfo{author}{\bibfnamefont{G.}~\bibnamefont{Hrkac}},
  \bibinfo{author}{\bibfnamefont{J.}~\bibnamefont{Dean}},
  \bibinfo{author}{\bibfnamefont{S.}~\bibnamefont{Bance}},
  \bibinfo{author}{\bibfnamefont{D.}~\bibnamefont{Suess}},
  \bibinfo{author}{\bibfnamefont{O.}~\bibnamefont{Ertl}},
  \bibinfo{author}{\bibfnamefont{F.}~\bibnamefont{Dorfbauer}},
  \bibnamefont{and} \bibinfo{author}{\bibfnamefont{J.}~\bibnamefont{Fidler}},
  \bibinfo{journal}{Applied Physics Letters} \textbf{\bibinfo{volume}{91}},
  \bibinfo{eid}{222502} (\bibinfo{year}{2007}).

\bibitem[{\citenamefont{Aharoni}(1960)}]{aharoni_1960}
\bibinfo{author}{\bibfnamefont{A.}~\bibnamefont{Aharoni}},
  \bibinfo{journal}{Phys. Rev.} \textbf{\bibinfo{volume}{119}},
  \bibinfo{pages}{127} (\bibinfo{year}{1960}).

\bibitem[{\citenamefont{Asti et~al.}(2004)\citenamefont{Asti, Solzi, Ghidini,
  and Neri}}]{asti_2004}
\bibinfo{author}{\bibfnamefont{G.}~\bibnamefont{Asti}},
  \bibinfo{author}{\bibfnamefont{M.}~\bibnamefont{Solzi}},
  \bibinfo{author}{\bibfnamefont{M.}~\bibnamefont{Ghidini}}, \bibnamefont{and}
  \bibinfo{author}{\bibfnamefont{F.~M.} \bibnamefont{Neri}},
  \bibinfo{journal}{Phys. Rev. B} \textbf{\bibinfo{volume}{69}},
  \bibinfo{pages}{174401} (\bibinfo{year}{2004}).

\bibitem[{\citenamefont{Asti et~al.}(2006)\citenamefont{Asti, Ghidini,
  Pellicelli, Pernechele, Solzi, Albertini, Casoli, Fabbrici, and
  Pareti}}]{asti_2006}
\bibinfo{author}{\bibfnamefont{G.}~\bibnamefont{Asti}},
  \bibinfo{author}{\bibfnamefont{M.}~\bibnamefont{Ghidini}},
  \bibinfo{author}{\bibfnamefont{R.}~\bibnamefont{Pellicelli}},
  \bibinfo{author}{\bibfnamefont{C.}~\bibnamefont{Pernechele}},
  \bibinfo{author}{\bibfnamefont{M.}~\bibnamefont{Solzi}},
  \bibinfo{author}{\bibfnamefont{F.}~\bibnamefont{Albertini}},
  \bibinfo{author}{\bibfnamefont{F.}~\bibnamefont{Casoli}},
  \bibinfo{author}{\bibfnamefont{S.}~\bibnamefont{Fabbrici}}, \bibnamefont{and}
  \bibinfo{author}{\bibfnamefont{L.}~\bibnamefont{Pareti}},
  \bibinfo{journal}{Phys. Rev. B} \textbf{\bibinfo{volume}{73}},
  \bibinfo{pages}{094406} (\bibinfo{year}{2006}).

\bibitem[{\citenamefont{Ghidini et~al.}(2007)\citenamefont{Ghidini, Asti,
  Pellicelli, Pernechele, and Solzi}}]{ghidini_2007}
\bibinfo{author}{\bibfnamefont{M.}~\bibnamefont{Ghidini}},
  \bibinfo{author}{\bibfnamefont{G.}~\bibnamefont{Asti}},
  \bibinfo{author}{\bibfnamefont{R.}~\bibnamefont{Pellicelli}},
  \bibinfo{author}{\bibfnamefont{C.}~\bibnamefont{Pernechele}},
  \bibnamefont{and} \bibinfo{author}{\bibfnamefont{M.}~\bibnamefont{Solzi}},
  \bibinfo{journal}{Journal of Magnetism and Magnetic Materials}
  \textbf{\bibinfo{volume}{316}}, \bibinfo{pages}{159 } (\bibinfo{year}{2007}).

\bibitem[{\citenamefont{Pellicelli et~al.}(2010)\citenamefont{Pellicelli,
  Solzi, Neu, H\"afner, Pernechele, and Ghidini}}]{pellicelli_2010}
\bibinfo{author}{\bibfnamefont{R.}~\bibnamefont{Pellicelli}},
  \bibinfo{author}{\bibfnamefont{M.}~\bibnamefont{Solzi}},
  \bibinfo{author}{\bibfnamefont{V.}~\bibnamefont{Neu}},
  \bibinfo{author}{\bibfnamefont{K.}~\bibnamefont{H\"afner}},
  \bibinfo{author}{\bibfnamefont{C.}~\bibnamefont{Pernechele}},
  \bibnamefont{and} \bibinfo{author}{\bibfnamefont{M.}~\bibnamefont{Ghidini}},
  \bibinfo{journal}{Phys. Rev. B} \textbf{\bibinfo{volume}{81}},
  \bibinfo{pages}{184430} (\bibinfo{year}{2010}).

\bibitem[{\citenamefont{Dobin and Richter}(2006)}]{dobin_2006}
\bibinfo{author}{\bibfnamefont{A.~Y.} \bibnamefont{Dobin}} \bibnamefont{and}
  \bibinfo{author}{\bibfnamefont{H.~J.} \bibnamefont{Richter}},
  \bibinfo{journal}{Applied Physics Letters} \textbf{\bibinfo{volume}{89}},
  \bibinfo{eid}{062512} (\bibinfo{year}{2006}).

\bibitem[{\citenamefont{Skomski and Coey}(1993)}]{skomski_1993}
\bibinfo{author}{\bibfnamefont{R.}~\bibnamefont{Skomski}} \bibnamefont{and}
  \bibinfo{author}{\bibfnamefont{J.~M.~D.} \bibnamefont{Coey}},
  \bibinfo{journal}{Phys. Rev. B} \textbf{\bibinfo{volume}{48}},
  \bibinfo{pages}{15812} (\bibinfo{year}{1993}).

\bibitem[{\citenamefont{Skomski}(1994)}]{skomski_1994}
\bibinfo{author}{\bibfnamefont{R.}~\bibnamefont{Skomski}},
  \bibinfo{journal}{Journal of Applied Physics} \textbf{\bibinfo{volume}{76}},
  \bibinfo{pages}{7059} (\bibinfo{year}{1994}).

\bibitem[{\citenamefont{Leineweber and Kronmuller}(1997)}]{leineweber_1997}
\bibinfo{author}{\bibfnamefont{T.}~\bibnamefont{Leineweber}} \bibnamefont{and}
  \bibinfo{author}{\bibfnamefont{H.}~\bibnamefont{Kronmuller}},
  \bibinfo{journal}{Journal of Magnetism and Magnetic Materials}
  \textbf{\bibinfo{volume}{176}}, \bibinfo{pages}{145} (\bibinfo{year}{1997}).

\bibitem[{\citenamefont{Kronmüller and Goll}(2002)}]{kronmuller_2002}
\bibinfo{author}{\bibfnamefont{H.}~\bibnamefont{Kronmüller}} \bibnamefont{and}
  \bibinfo{author}{\bibfnamefont{D.}~\bibnamefont{Goll}},
  \bibinfo{journal}{Physica B: Condensed Matter}
  \textbf{\bibinfo{volume}{319}}, \bibinfo{pages}{122 } (\bibinfo{year}{2002}).

\bibitem[{\citenamefont{Guslienko et~al.}(2004)\citenamefont{Guslienko,
  Chubykalo-Fesenko, Mryasov, Chantrell, and Weller}}]{guslienko_2004}
\bibinfo{author}{\bibfnamefont{K.~Y.} \bibnamefont{Guslienko}},
  \bibinfo{author}{\bibfnamefont{O.}~\bibnamefont{Chubykalo-Fesenko}},
  \bibinfo{author}{\bibfnamefont{O.}~\bibnamefont{Mryasov}},
  \bibinfo{author}{\bibfnamefont{R.}~\bibnamefont{Chantrell}},
  \bibnamefont{and} \bibinfo{author}{\bibfnamefont{D.}~\bibnamefont{Weller}},
  \bibinfo{journal}{Phys. Rev. B} \textbf{\bibinfo{volume}{70}},
  \bibinfo{pages}{104405} (\bibinfo{year}{2004}).

\bibitem[{\citenamefont{Casoli et~al.}(2008)\citenamefont{Casoli, Albertini,
  Nasi, Fabbrici, Cabassi, Bolzoni, and Bocchi}}]{casoli_2008}
\bibinfo{author}{\bibfnamefont{F.}~\bibnamefont{Casoli}},
  \bibinfo{author}{\bibfnamefont{F.}~\bibnamefont{Albertini}},
  \bibinfo{author}{\bibfnamefont{L.}~\bibnamefont{Nasi}},
  \bibinfo{author}{\bibfnamefont{S.}~\bibnamefont{Fabbrici}},
  \bibinfo{author}{\bibfnamefont{R.}~\bibnamefont{Cabassi}},
  \bibinfo{author}{\bibfnamefont{F.}~\bibnamefont{Bolzoni}}, \bibnamefont{and}
  \bibinfo{author}{\bibfnamefont{C.}~\bibnamefont{Bocchi}},
  \bibinfo{journal}{Applied Physics Letters} \textbf{\bibinfo{volume}{92}},
  \bibinfo{eid}{142506} (\bibinfo{year}{2008}).

\bibitem[{\citenamefont{Tsai et~al.}(2010)\citenamefont{Tsai, Tzeng, and
  Liu}}]{tsai_2010}
\bibinfo{author}{\bibfnamefont{J.-L.} \bibnamefont{Tsai}},
  \bibinfo{author}{\bibfnamefont{H.-T.} \bibnamefont{Tzeng}}, \bibnamefont{and}
  \bibinfo{author}{\bibfnamefont{B.-F.} \bibnamefont{Liu}},
  \bibinfo{journal}{Thin Solid Films} \textbf{\bibinfo{volume}{518}},
  \bibinfo{pages}{7271 } (\bibinfo{year}{2010}).

\bibitem[{\citenamefont{Scholz et~al.}(2003)\citenamefont{Scholz, Fidler,
  Schrefl, Suess, Dittrich, Forster, and Tsiantos}}]{magpar_2003}
\bibinfo{author}{\bibfnamefont{W.}~\bibnamefont{Scholz}},
  \bibinfo{author}{\bibfnamefont{J.}~\bibnamefont{Fidler}},
  \bibinfo{author}{\bibfnamefont{T.}~\bibnamefont{Schrefl}},
  \bibinfo{author}{\bibfnamefont{D.}~\bibnamefont{Suess}},
  \bibinfo{author}{\bibfnamefont{R.}~\bibnamefont{Dittrich}},
  \bibinfo{author}{\bibfnamefont{H.}~\bibnamefont{Forster}}, \bibnamefont{and}
  \bibinfo{author}{\bibfnamefont{V.}~\bibnamefont{Tsiantos}},
  \bibinfo{journal}{Comp. Mat. Sci.} \textbf{\bibinfo{volume}{28}},
  \bibinfo{pages}{366} (\bibinfo{year}{2003}).

\bibitem[{\citenamefont{Dobin and Richter}(2007)}]{dobin_2007}
\bibinfo{author}{\bibfnamefont{A.~Y.} \bibnamefont{Dobin}} \bibnamefont{and}
  \bibinfo{author}{\bibfnamefont{H.~J.} \bibnamefont{Richter}},
  \bibinfo{journal}{Journal of Applied Physics} \textbf{\bibinfo{volume}{101}},
  \bibinfo{eid}{09K108} (\bibinfo{year}{2007}).

\end{thebibliography}

\eject
 
\newpage                                         
\clearpage                                         
\clearpage                                         
\clearpage                                         
%
\clearpage                                         
\end{document}